\documentclass[prb,aps,twocolumn,superscriptaddress,showpacs]{revtex4}

\usepackage{graphicx}
\usepackage{verbatim}

\begin{document}

\title{Soft phonon columns on the edge of the Brillouin zone in the relaxor PbMg$_{1/3}$Nb$_{2/3}$O$_{3}$}

\author{I.P. Swainson}
\affiliation{{National Research Council, Chalk River, Ontario,
Canada  KOJ 1JO}}

\author{C. Stock}
\affiliation{ISIS Facility, Rutherford Appleton Laboratory, Didcot,
Oxon, United Kingdom  OX11 0QX}

\author{P. M. Gehring}
\affiliation{NIST Center for Neutron Research, National Institute of
Standards and Technology, Gaithersburg, Maryland  20899}

\author{Guangyong Xu}
\affiliation{Condensed Matter Physics and Materials Science
Department, Brookhaven National Laboratory, Upton, New York  11973}

\author{K. Hirota}
\affiliation{Institute for Solid State Physics, University of Tokyo,
Kashiwa Japan  277-8581}

\author{Y. Qiu}
\affiliation{NIST Center for Neutron Research, National Institute of
Standards and Technology, Gaithersburg, Maryland  20899}
\affiliation{Department of Materials Science and Engineering,
University of Maryland, College Park, Maryland  20742}

\author{H. Luo}
\affiliation{Shanghai Institute of Ceramics, Chinese Academy of
Sciences, Shanghai, China  201800}

\author{X. Zhao}
\affiliation{Shanghai Institute of Ceramics, Chinese Academy of
Sciences, Shanghai, China  201800}

\author{J.-F. Li}
\affiliation{Department of Materials Science and Engineering,
Virginia Tech., Blacksburg, Virginia 24061}

\author{D. Viehland}
\affiliation{Department of Materials Science and Engineering,
Virginia Tech., Blacksburg, Virginia 24061}

\date{\today}

\begin{abstract}

We report lattice dynamical measurements, made using neutron inelastic scattering methods, of the relaxor perovskite PbMg$_{1/3}$Nb$_{2/3}$O$_{3}$ (PMN) at momentum transfers near the edge of the Brillouin zone.  Unusual``columns'' of phonon scattering that are localized in momentum, but extended in energy, are seen at both high-symmetry points along the zone edge: $\vec{Q}_R=\left\{\frac{1}{2}, \frac{1}{2}, \frac{1}{2}\right\}$ and $\vec{Q}_M=\left\{\frac{1}{2},\frac{1}{2},0\right\}$.  These columns soften at $\sim$400\,K which is similar to the onset temperature of the zone-center diffuse scattering, indicating a competition between ferroelectric and antiferroelectric distortions. We propose a model for the atomic displacements associated with these phonon modes that is based on a combination of structure factors and group theoretical analysis.  This analysis suggests that the scattering is not from tilt modes (rotational modes of oxygen octahedra), but from zone-boundary optic modes that are associated with the displacement of Pb$^{2+}$ and O$^{2-}$ ions. Whereas similar columns of scattering have been reported in metallic and (less commonly) molecular systems, they are unusual in insulating materials, particularly in ferroelectrics; therefore, the physical origin of this inelastic feature in PMN is unknown.  We speculate that the underlying disorder contributes to this unique anomaly.  

\end{abstract}

\maketitle

\section{Introduction}

The lead-based, relaxor ferroelectrics are chemically disordered compounds that have attracted considerable scientific interest due to their remarkable structural, dynamical, and electromechanical properties.~\cite{Ye98:81,Bokov_rev,Hirota06:11} PbMg$_{1/3}$Nb$_{2/3}$O$_{3}$ (PMN) and PbZn$_{1/3}$Nb$_{2/3}$O$_{3}$ (PZN) are two of the most widely studied compounds, and both display the broad, frequency-dependent peak in the dielectric response as a function of temperature that is characteristic of relaxors.~\cite{Viehland90:68} Surprisingly, even though both possess unusually large dielectric susceptibilities ($\epsilon \sim$ 25,000), neither PMN nor PZN exhibits a well-defined structural transition, and thus no ferroelectricity, in zero electric field.~\cite{Bonneau89:24,Mathan91:3,Xu04:70} When diluted with sufficient amounts of PbTiO$_{3}$ (PT), which is a conventional ferroelectric, a long-range ferroelectric ground state is established and the piezoelectric coefficients grow to record-setting levels, making these materials prime candidates for industrial applications. But beyond a certain concentration of PbTiO$_{3}$, which defines the Morphotropic Phase Boundary (MPB) that separates the low-concentration monoclinic phase from the high-concentration tetragonal phase, the piezoelectric coefficients drop precipitously.~\cite{Park97:82,Liu99:85,Kiat02:65} Whereas different ideas have been proposed to explain these effects, no consensus exists on the origin of the exceptional piezoelectric properties.

The static properties of relaxors are highly unusual.  A structural distortion was observed in single-crystal PZN using low-energy x-ray diffraction techniques,~\cite{Lebon} but this was subsequently shown to be confined to a near-surface region of the crystal by another diffraction study that employed higher-energy x-rays.~\cite{Xu03:67} This localized structural distortion, known as the ``skin effect,'' is enhanced by strong electric fields and has been confirmed using neutron strain-scanning techniques in other PZN- and PMN-based compounds.~\cite{Conlon04:70,Xu06:79,Stock07:unpub,Gehring04:16} Despite the absence of a bulk ferroelectric ground state in either PMN or PZN, evidence of static, short-range, polar correlations has been inferred from measurements of the optic index of refraction~\cite{Burns83:48} and by the presence of temperature-dependent diffuse scattering that is centered on the Bragg peaks.~\cite{Naberezhnov} The refraction measurements have been interpreted in terms of local, nanometer-scale regions of randomly oriented ferroelectric order, often termed polar
nanoregions (PNR), which form below the Burns temperature, $T_B$, which is nominally 620\,K in PMN\@.~\cite{Burns83:48} Various models of the diffuse scattering and low-energy phonons have been proposed; however, all models associate the diffuse scattering with the formation of polar
nanoregions in PMN and PZN, because both appear at the temperature $T_B$\@.~\cite{You97:79,Hirota02:65,Welberry05:38,Vug06:73} For PT-concentrations beyond the MPB, no diffuse scattering is observed. As this coincides with the precipitous drop in the size of the piezoelectric coefficients, the presence of substantial elastic diffuse scattering near the zone center is associated with the relaxor effect.~\cite{Hiraka04:70,Matsuura06:74}

Whereas this diffuse scattering represents a definitive, \textit{static}
signature of relaxors, extensive research has yet to identify a
well-defined \textit{dynamical} property that is uniquely associated with
relaxor behavior.  Being chemically disordered solids, the lattice
dynamics of PMN and PZN are quite complex.  Nevertheless, important
similarities exist between the two compounds.  Like the parent
material PbTiO$_3$, which is a soft-mode (displacive) ferroelectric,
both PMN and PZN exhibit phonon spectra near the zone center that
are dominated by a soft transverse optic (TO) mode that reaches a
minimum frequency on cooling and then hardens at lower
temperatures.~\cite{Gehring01:87,Naberezhnov,Wakimoto02:65,Stock04:69,Gvasaliya04:69,Gvasaliya05:17}
Concurrent with the decrease in frequency, the linewidth of this
mode broadens in energy until it becomes heavily damped, but only
for wave vectors near the zone center.  This anomalous temperature-
and wave vector-dependent damping produces a false, dispersion-like
feature known as the ``waterfall,'' which has been observed in PMN,
PZN, and other relaxor compounds.~\cite{Gehring_Aspen,Gehring_pzn}
Early studies speculated that the phonon damping resulted from a
strong interaction between the soft mode and the PNR\@.~\cite{Gehring_pzn8pt} 
Since then, however, the origin of the
waterfall effect has been a topic of considerable
debate.~\cite{Hlinka03:91,Hlinka08:81,Vak02:66} There is now strong
evidence that refutes the idea that the waterfall effect is caused
by TO-PNR coupling.  Low-frequency phonon studies of PMN mixed
with 60\%\ PbTiO$_3$ (PMN-60PT), a composition that is not a relaxor
and that displays a long-range, first-order, cubic--tetragonal phase
transition, have found a waterfall feature very similar to that seen
in PMN\@.~\cite{Stock06:73}  Attempts to find a waterfall 
in the K(Ta$_{1-x}$Li$_{x}$)O$_{3}$ relaxor system have not been successful.~\cite{Wakimoto06:74}  
Thus, the heavily damped soft mode is not
a defining characteristic of relaxors, and its existence does not
require the presence of PNR\@.  Instead, the waterfall feature probably
results from another form of strong disorder common to both relaxor
and non-relaxor ferroelectrics.  A dynamic feature that is very
probably associated with the PNR (i.e., with the diffuse scattering), and 
thus the
relaxor phenomenon, is the strong damping of the low-energy
transverse acoustic phonons.  This damping was recently reported to
have a strong dependence on the electric field and is absent in PMN-$x$PT
compositions that lie beyond the MPB\@.~\cite{Stock05:74,Xu08:unpub}  

No signature of the relaxor phase has yet been identified from studies of the zone-center soft optic mode, although it was recently suggested that the MPB begins at that PT concentration where the structural transition temperature matches the temperature at which the zone-center soft phonon mode reaches a minimum.~\cite{Cao08:78} The existence of the two temperature scales in the relaxor phase has been explained by random field theories.~\cite{Stock04:69,Fisch03:67,Westphal92:68} Finding such a feature in PMN is extremely important because, as is the case with soft modes in conventional ferroelectrics, a unique dynamical signature can provide information on the underlying structural distortions that are specific to the relaxor state.  In view of the similarities of the zone-center soft mode dynamics throughout the relaxor-ferroelectric phase diagram, such as the waterfall effect, we believe it is necessary to search other regions of the Brillouin zone for a dynamic signature that is unique to relaxors.

\begin{figure}[t]
\includegraphics[width=2.5in]{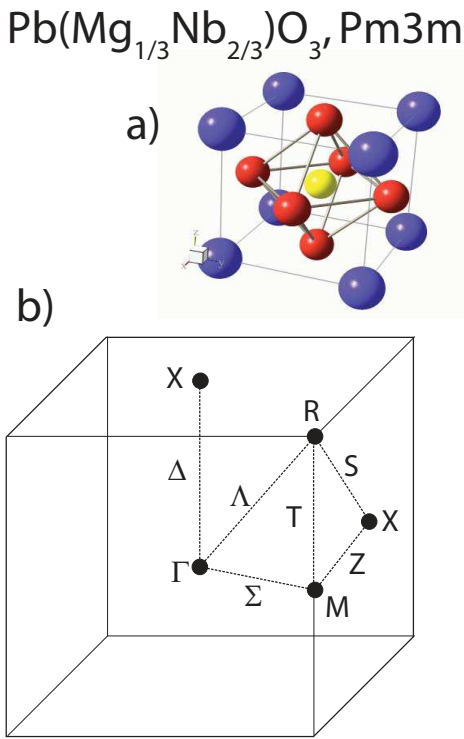}
\caption{(Color online) a) The undistorted, average cubic unit cell of
PMN is depicted with the Mg$_{1/3}$Nb$_{2/3}$O$_3$ octahedron in the
center and Pb cations at the corners. The blue, red, and yellow
spheres represent Pb$^{2+}$, O$^{2-}$, and Mg$^{2+}$/Nb$^{5+}$ ions,
respectively.  b) The Brillouin zone of a primitive cubic lattice is
shown using the labels of Miller and Love~\cite{Miller-Love} to
denote the high-symmetry points and lines.  $M$, $R$, and $X$
represent $\vec{Q}=\left\{\frac{1}{2},\frac{1}{2},0\right\}$,
$\left\{\frac{1}{2},\frac{1}{2},\frac{1}{2}\right\}$, and
$\left\{\frac{1}{2},0,0\right\}$ reciprocal lattice vectors,
respectively.}
 \label{fig:struct-BZ}
\end{figure}

In comparison to the zone center, little experimental attention has been paid to the
zone boundary of relaxor ferroelectrics, but intriguing temperature-dependent superlattice peaks have been observed in PMN using TEM and x-ray
diffraction techniques at the reciprocal lattice vectors $(\frac{1}{2},\frac{1}{2},0)$ and
$(\frac{1}{2},\frac{1}{2},\frac{1}{2})$.~\cite{Hilton,Vakhrushev_ZB,Tkachuk,Tkachuk2}
Being zone boundary locations these would seem unlikely to be
associated with the relaxor transition, as any softening would be
indicative of antiferroelectric, rather than ferroelectric,
distortions.  However, it has been suggested theoretically that both
ferroelectric and antiferroelectric distortions can coexist, and
such a concept has been applied to the case of SrTiO$_{3}$ in 
Ref.~\onlinecite{Zong95:74}\@. The average, long-range structure of the
cubic phase of PMN is illustrated in Fig.~\ref{fig:struct-BZ}, as
are the high-symmetry positions of the associated Brillouin zone.
There are two high-symmetry points along the edge of the Brillouin
zone: $M=\left\{\frac{1}{2},\frac{1}{2},0\right\}$, and
$R=\left\{\frac{1}{2},\frac{1}{2},\frac{1}{2}\right\}$. In this
paper we report the presence of soft phonons at both of these
high-symmetry points in PMN\@.  Unlike the highly damped TO phonons
measured near the zone center, the zone-boundary points display
vertical columns of phonon scattering that are especially strong at
$M$, and these appear at a temperature that coincides with the
appearance of diffuse scattering at the zone center.  Such columns
of inelastic scattering are very unusual in insulators, and their ultimate
origin remains unknown at this time.  We focus only on
the soft phonons in this study and do not deal directly with the
full complexity of the elastic diffuse scattering near the zone
boundaries reported elsewhere.~\cite{Gosula00:61}

This paper is divided into three sections.  The first section
provides the experiment details of the measurements conducted at the
respective neutron scattering facilities located at the National
Institute of Standards and Technology Center for Neutron Research
(NCNR) and Canadian Neutron Beam Centre at the Chalk River Laboratories\@. The second section deals with the static properties of PMN, in particular the temperature dependence of the elastic scattering measured at the zone boundary.  The third section describes the
inelastic measurements showing the presence of a
temperature-dependent column of scattering that is associated with
the temperature-dependent elastic diffuse scattering. We compare
these results to other systems and, in particular, to PMN-60PT,
which is not a relaxor, but undergoes a long-range, cooperative
cubic--tetragonal transition, similar to that occuring in the parent material
PbTiO$_3$\@.~\cite{Shirane70:2a}

\section{Experimental Details}

The measurements described in this paper are based on two single crystals
with volumes of 3\,cc and 9\,cc.  Both samples were grown using the
Bridgeman technique described elsewhere.~\cite{Luo00:39}  The large
9\,cc sample is the same crystal that was used in a previous phonon
study.~\cite{Stock05:74}  Both samples have a room temperature
lattice constant $a = 4.04$\,\AA; thus one reciprocal lattice unit
(1\,rlu) equals $2\pi/a = 1.56$\,{\AA}$^{-1}$. 

Neutron inelastic scattering experiments were carried out on the
time-of-flight Disk Chopper Spectrometer (DCS, NCNR), the SPINS cold
neutron triple-axis spectrometer (NCNR) and the C5 and N5 thermal neutron
triple-axis spectrometers (Chalk River)\@.  The experiments 
conducted at both the N5
and C5 thermal triple-axis spectrometers used a cryofurnace
to vary and control the sample temperature between 10\,K and 600\,K\@.
Experiments conducted on the DCS and SPINS instruments employed a
high-temperature displex over a similar temperature range. To
prevent any possible decomposition, the sample was never heated
beyond 600\,K\@ and neither sample was ever exposed to an electric field at any temperature.  
Next we describe the experimental setups used for
the various neutron scattering measurements on each instrument.  All 
uncertainties (error bars) shown in the figures in
this paper correspond to the square root of the measured
neutron intensity.

\subsection{Triple-axis Spectrometers (C5, N5, and SPINS)}

Measurements on the C5 thermal triple-axis spectrometer were made with a variable-focusing 
(002) pyrolytic graphite (PG) monochromator and a flat PG
(002) analyzer.  The horizontal angular beam collimation sequence
was set to either 12'-33'-$S$-29'-72' or 33'(open)-33'-$S$-29'-72'
($S =$ Sample), and a fixed, final neutron energy $\rm E_f
=14.6$\,meV was used to define the energy transfer as $\hbar \omega
= {\rm E}_i-\rm{E}_f$. A PG filter was placed after the sample to
remove neutrons scattering from higher order reflections of the
monochromator and a sapphire filter was placed before the
monochromator to remove high-energy, ``fast'' neutrons. The
instrumental elastic energy resolution, defined as the full width at
half maximum (FWHM) of the elastic ($\hbar \omega=0$) peak, was
$\sim$0.9\,meV for both collimation sequences. The sample was
aligned in the (HK0), (HHL), and the (H3HL) scattering planes. The
(HK0) plane allows a complete investigation of the zone center
acoustic phonons as described in Ref.~\onlinecite{Stock05:74}.
Through the use of the lower symmetry (H3HL) scattering plane, we are
sensitive to the zone-boundary acoustic modes, which are 
tilt (or rotary modes) as outlined in
Ref.~\onlinecite{Lynn78:27}.  

Experiments conducted on N5 utilized both a flat PG (002)
monochromator and analyzer.  The final (scattered) neutron energy
was fixed to $\rm E_f=14.8$\,meV, the collimation was set at
30'(open)-40'-$S$-40'-open, and a PG filter was placed after the
sample.  The instrumental elastic energy resolution was 1.3\,meV
FWHM\@. For these experiments the sample was aligned in the (HHL)
scattering plane.  We note that all of the inelastic data presented
in this paper taken on triple-axis spectrometers have been corrected
for contamination of the incident beam monitor as described
elsewhere.~\cite{Shirane:book}

Measurements on SPINS used a
vertically-focused PG (002) monochromator and a flat PG (002)
analyzer. The final neutron energy was fixed to $\rm E_f=4.5$\,meV
and the horizontal beam collimations were set to
(guide)-80'-80'-(open) resulting in an instrumental elastic energy
resolution of 0.25\,meV FWHM\@. A liquid-nitrogen-cooled beryllium
filter was placed after the sample to remove higher-order neutron
contamination.

\subsection{Disk Chopper Spectrometer (DCS)}

\begin{figure}[t]
\includegraphics[width=3.7in]{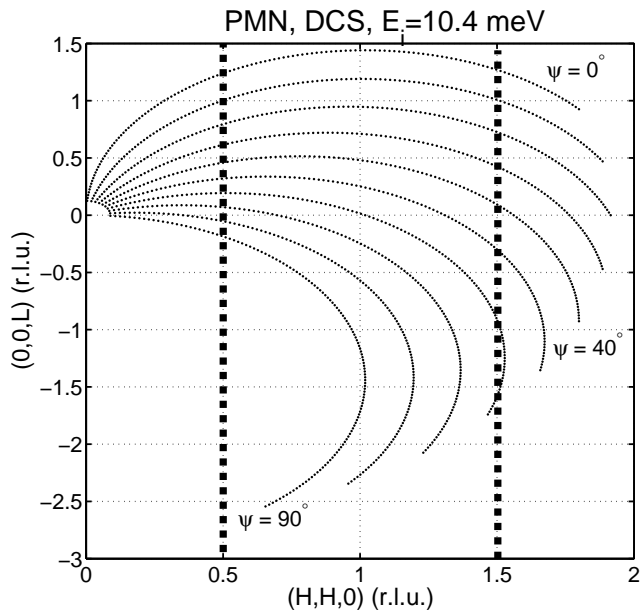}
\caption{The detector trajectories, at zero energy transfer, for measurements on DCS with E$_{i}$ = 10.4\,meV in the (HHL) scattering planes.  The trajectories are plotted from $\psi=10^{\circ}$ to $90^{\circ}$ with $10^{\circ}$ spacing for clarity.  The vertical dashed lines at ${\rm H} = 0.5$ and 1.5 indicate the $T$ lines along the Brillouin zone edge (Fig.~\protect\ref{fig:struct-BZ})\@.}
 \label{fig:det_pos}
\end{figure}

Measurements on the Disk Chopper Spectrometer (DCS) were made at
$\sim$40 different successive crystal orientations, spaced
$2^{\circ}$ apart about a vertical axis, at 300\,K and 600\,K\@.  The
DCS instrument consists of 325 detectors in the horizontal
scattering plane (or 913 including the out-of-plane detectors) with
active dimensions in and normal to the scattering plane of
$\sim$3.1\,cm and 40\,cm, respectively.  For all data presented
here, only the central detector bank was used. The detectors are
located 4\,m from the sample at scattering angles ranging from
$2\theta=5^{\circ}$ to $140^{\circ}$. Further details of the
instrument can be found elsewhere.~\cite{Copley03:292}  The
incident energy was fixed at 6.0\,meV and 10.4\,meV for measurements
made in the (HK0) and (HHL) scattering planes, respectively,
resulting in instrumental elastic energy resolutions of 0.25\,meV
and 0.6\,meV FWHM\@.  The data were binned and visualized using the
DCS MSLICE program.~\cite{Qiu:mslice}  To obtain the energy
positions and linewidths throughout the Brillouin zone, we fit the
data to a linear combination of simple damped harmonic oscillators.
This lineshape is described in detail elsewhere (Ref.~\onlinecite{Stock05:74,Stock06:73,Cowley73:6}).

The DCS experiments were conducted on the neutron energy-gain side,
where $\hbar \omega = \rm E_i- \rm E_f$ is negative, whereas the
bulk of our triple-axis work was conducted on the neutron
energy-loss side where $\hbar \omega$ is positive.  The intensity of
the scattering on the energy-gain side is related to that on the
energy-loss side through detailed balance.~\cite{Shirane:book}
Therefore, throughout this paper we retain the convention that
energy transfer is defined by $\hbar \omega=E_i-E_f$ and
indicate the energy transfer to be negative for data from DCS and
positive for the bulk of our triple-axis work. This experimental
setup provided good energy resolution near the elastic position
(near $\hbar \omega=0$), which allowed us to measure the details of
the low-energy acoustic and optic modes, and coarser energy
resolution at larger negative energy transfers, where contamination
from elastic scattering is not a concern.

The use of a chopper instrument may seem like an unusual choice for study phonons which have dispersion in energy in all three directions of momentum.  The use of DCS is different from that of a triple-axis as there are many more detectors whose position and scattering angle are fixed.  By measuring the intensity at each detector as a function of time (and hence as a function of energy transfer), a trajectory in $\vec{Q}$-$E$ space is obtained.  To obtain a complete map of reciprocal space as a function of energy transfer, many successive crystal orientations need to be obtained.  A summary of the momentum space coverage at a series of crystal orentations is illustrated Fig. \ref{fig:det_pos}.  The angle $\psi$ is defined as the angle between the [001] axis of the crystal and the incident beam.  Each line in the figure corresponds to a particular crystal orientation and it can be seen that by rotating the sample orientation with respect to the incident beam, an entire map of $\vec{Q}$-$E$  space can be obtained.  Throughout this paper, we have symmetrized the data around ${\rm L} = 0$ and plotted it in terms of positive values of L\@. 

\section{Elastic Scattering}

\begin{figure}[t]
\includegraphics[width=3.25in]{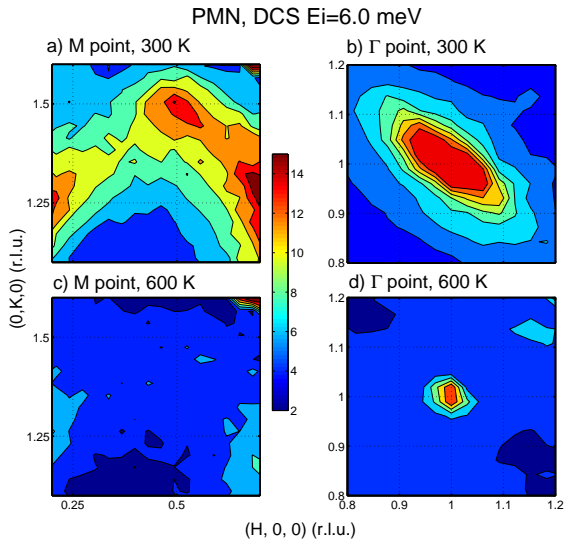}
\caption{Color contour plots of the elastic scattering at 300\,K and
600\,K near $M$ (panels a) and c)) and near $\Gamma$ (panels b)
and d)). The data have been integrated from $\hbar \omega=-1$\,meV to
1\,meV and plotted in reciprocal lattice units (r.l.u.). These data
were taken on the DCS instrument with the crystal aligned in the
(HK0) scattering plane.}
 \label{fig:M-Gamma}
\end{figure}

We first examine the elastic scattering measured near all three
high-symmetry points on the zone boundary ($M$, $X$, and $R$, as
defined in Fig.~\ref{fig:struct-BZ}). Figure~\ref{fig:M-Gamma}
illustrates contour plots of the elastic intensity measured with the
DCS instrument in the (HK0) scattering plane in the vicinity of the
reciprocal lattice vectors  $\vec{Q}_M=(\frac{1}{2},\frac{3}{2},0)$
and $\vec{Q}_\Gamma=(1,1,0)$.  At 600\,K there is no observable
scattering around $M$, and the $\Gamma$-point is characterized by a
sharp Bragg peak.  At 300\,K strong diffuse scattering is seen near
the $\Gamma$-point with a lineshape in $q$ (the reduced wave vector,
which is measured relative to the zone center in reciprocal lattice units) that is consistent
with previous measurements.~\cite{Hiraka04:70} A broad peak is also
observed at the $M$-point (Fig.~\ref{fig:M-Gamma}$a$) indicating a
second type of structural instability.  The fact that the scattering
is not resolution-limited in $q$ at the $M$-point indicates that any
static correlations are short-range and thus do not correspond to a
well-defined structural distortion: long-range correlations would
result in a resolution-limited Bragg peak like that shown in
Fig.~\ref{fig:M-Gamma}$d$).

\begin{figure}[t]
\includegraphics[width=3.0in]{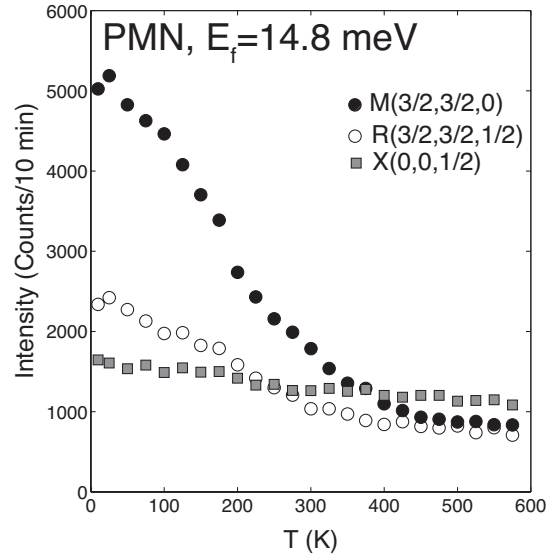}
\caption{The temperature dependence of the elastic scattering
measured at $\vec{Q}_M=(\frac{3}{2}, \frac{3}{2}, 0)$, $\vec{Q}_R =
(\frac{3}{2}, \frac{3}{2}, \frac{1}{2})$, and $\vec{Q}_X =(0, 0,
\frac{1}{2})$. Whereas a strong temperature dependence is observed
near the $M$- and $R$-points, little temperature dependence is
observed near the $X$-point.  These data were taken on the N5
triple-axis spectrometer.} \label{fig:zb_temp}
\end{figure}

The temperature dependence of the zone boundary elastic scattering
was studied on the N5 triple-axis spectrometer and is illustrated in
Fig.~\ref{fig:zb_temp}.  The intensity at the $X$-point increases
with decreasing temperature, but this growth matches the change in
transmission through the crystal (measured with the straight through
beam) and is otherwise featureless; thus it does not represent any
intrinsic structural change. By contrast, the intensities measured
at both the $M$- and $R$-points show a strong temperature dependence
and a clear upturn near 400\,K\@.  These data are consistent with the
available temperature-dependent x-ray data of Gosula \textit{et al.}\
(Ref.~\onlinecite{Gosula00:61}), except that we have extended the
temperature range up to 600\,K and show that the onset is near
400\,K\@.  The enhancement of the elastic scattering as a function of temperature
at the $M$- and $R$-points is suggestive of incipient structural
instabilities; however, it is also relatively smooth and is not
associated with a sharp transition to a long-range, structurally
distorted phase, which would result in a new Bragg peak.  As 
the scattering is diffuse, these data infer
the onset and subsequent growth of short-range, antiferrodistortive
correlations below 400\,K\@. This finding is significant
because 400\,K is the same temperature at which the zone-center TO
mode reaches a minimum in energy~\cite{Wakimoto02:65} and at which
the strong, zone-center diffuse scattering, which is associated with
the formation of static polar nanoregions, first appears.~\cite{Hiraka04:70}  Indeed, that the onset of zone center diffuse scattering occurs at 400 K in PMN motivated a very recent neutron scattering study of the Burns temperature using extremely sharp energy resolution and resulted in a reassessed value of T$_{B}$ = 420 K $\pm$ 20 K, roughly 200 K below the value identified by Burns and Dacol in 1983.~\cite{Gehring_TB}

Zone-boundary elastic scattering has been observed with
neutrons and transmission electron spectroscopy in
PbZr$_{0.52}$Ti$_{0.48}$O$_3$ by Noheda \textit{et al.}, however,
these superlattice peaks were determined to be from a minority
domain with antiphase rotations of the oxygen octahedra about
[111].~\cite{Noheda02:66} X-ray studies of both PMN and PSN
(PbSc$_{1/2}$Nb$_{1/2}$O$_{3}$) have also found evidence of diffuse
scattering at the $M$- and
$R$-points.~\cite{Takesue99:82,Gosula00:61}  PSN is unique as the
perovskite $B$ site is occupied equally by Sc$^{3+}$ and Nb$^{5+}$
and, therefore, ordering on this site may be expected to induce at
least short-range correlations at either the $M$- or $R$-points\@. The
$\frac{1}{3} : \frac{2}{3}$ stoichiometry on the $B$ site of PMN
does not allow such ordering, which means that short-range
correlations at either $M$ or $R$ are not necessarily expected.
Consequently, it is important to search for any phonon anomaly at the
$M$- and $R$-points to determine the symmetry of the distortion, as
well as to determine whether the elastic scattering originates in
the bulk, and thus is an intrinsic effect, or whether it is limited
to the surface or ``skin" region and is influenced by strain or local
compositional inhomogeneities.

\section{Inelastic Scattering}

\begin{figure}[t]
\includegraphics[width=3.5in]{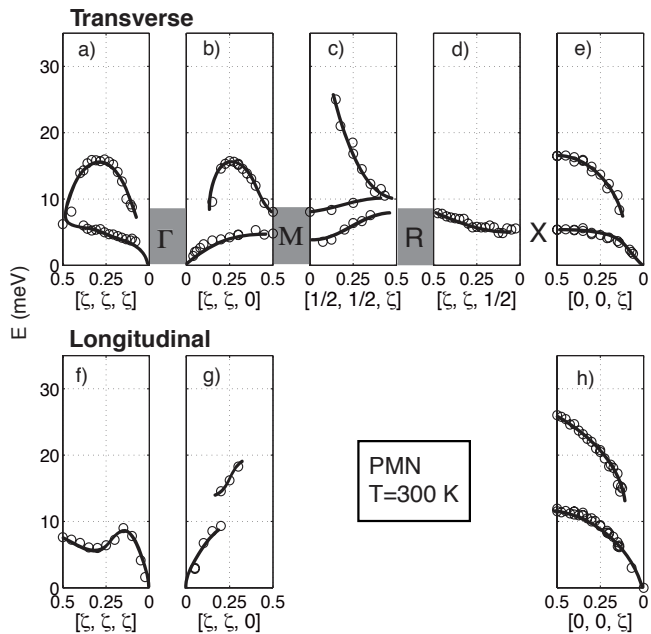}
\caption{A compilation of the room temperature ($T=300$\,K) phonon
dispersion curves in PMN measured along high symmetry directions
within the Brillouin zone. The locations of the incipient soft modes
are denoted by grey columns: $M$ (zone boundary), $\Gamma$ (zone
center), and $R$ (zone boundary). These data were obtained from
experiments conducted on the DCS chopper instrument, and the N5, C5,
and SPINS triple-axis spectrometers.  Panels a)-e) display the
transverse phonon dispersions and panels f)-h) show the
longitudinal dispersions.  The solid lines are guides to the eye.}
\label{fig:dispersion}
\end{figure}

Phonon dispersion measured below 30 meV, for both transverse
and longitudinal polarizations, are plotted in
Fig.~\ref{fig:dispersion}. These data are based on the experiments
described in the previous section and represent a compilation of
triple-axis and time-of-flight measurements.  Previous experiments
have focused on phonon branches that
include the $\Gamma$-point as well as on the temperature dependence
of the soft transverse optic and acoustic modes.  The data for those
measurements are summarized at room temperature in panels a), b),
and e) for modes propagating along the [111], [110], and [001]
directions, respectively.  Whereas we will not discuss the longitudinal
scattering in detail in this paper, we summarize our current data
set to define the energy scale of the phonon scattering in PMN in
panels f), g), and h): the energy scale is similar to that
reported for PbTiO$_3$\@.

The dispersion curves reveal several possible incipient soft mode
positions, these being where the optic modes reach a minimum energy
as a function of reduced wave vector.  Aside from the soft TO
mode at the $\Gamma$-point, which has been studied in detail
by several groups, optic mode minima are seen at the $M$- and
$R$-points; these minima correspond to the locations where strong,
temperature-dependent elastic scattering was observed, as described
in the previous section.   In this paper we concentrate on the phonons
and the dispersion curves along the $T$-line, which is
the zone-boundary edge that links the high-symmetry $M$- and
$R$-points (see Fig.\ \ref{fig:dispersion} c)).  We next describe
the temperature and wave vector dependence of the soft modes, and we assign these
to specific optic modes based on our measurements conducted with
time-of-flight and triple-axis spectrometers.  More importantly, we show that these
soft modes cannot be the tilt instabilities in cubic
perovskites, which are also associated with phonon instabilities at
these same points along the edge of the Brillouin zone.

\subsection{DCS Time-of-Flight Spectra}

\begin{figure}[t]
\includegraphics[width=3.5in]{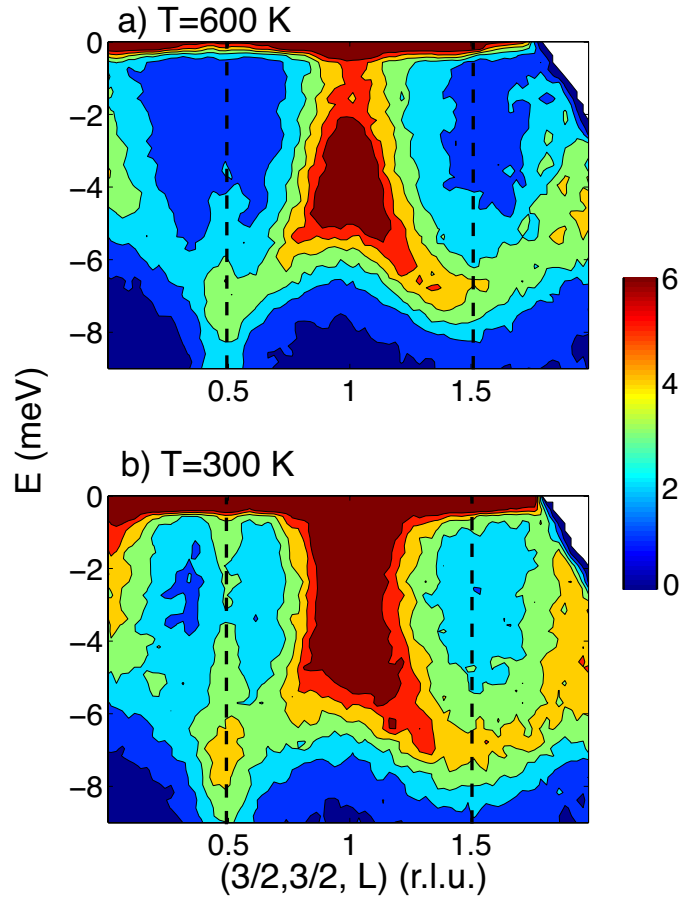}
\caption{Inelastic scattering measured at 300\,K and 600\,K along
the $T$-edge of the Brillouin zone $(\vec{Q}_T = \frac{3}{2},
\frac{3}{2}, {\rm L})$, where ${\rm L}=1$ is an $M$-point, and ${\rm L} = \frac{1}{2},
\frac{3}{2}$ are $R$-points, indicated by
the vertical dashed lines. These data are integrated over
($\delta$H, $\delta$H, 0) = ($\pm$ 0.05, $\pm$ 0.05, 0) in r.l.u.
and were taken on the DCS instrument with E$_{i}$=10.4\,meV\@.  The
negative energy transfer on the $y$-axis corresponds to an energy
gain of the neutron as explained in the text.  The intensity has been normalized by the Bose 
factor to put the intensities on an equal scale.} \label{fig:DCS}
\end{figure}

We summarize the unusual zone-boundary phonon scattering in
Fig.~\ref{fig:DCS}, which displays DCS spectra reconstructed along
the $T$-line $(\frac{3}{2}, \frac{3}{2}, {\rm L})$ with $\vec{Q}_M
=(\frac{3}{2}, \frac{3}{2}, 1)$ and $\vec{Q}_R =(\frac{3}{2},
\frac{3}{2}, \frac{1}{2})$ and $(\frac{3}{2}, \frac{3}{2},
\frac{3}{2})$ at 600\,K and 300\,K\@. These color contour plots illustrate
three key features regarding the nature of the lattice dynamics near
the zone boundaries.

First, a soft mode is evident at the $M$-point in PMN given the
substantial increase in spectral weight that occurs at low energies
upon cooling from 600\,K to 300\,K, which results in the column of
scattering shown at $\vec{Q}_M =(\frac{3}{2}, \frac{3}{2}, 1)$ in
Fig.~\ref{fig:DCS}. At 600\,K (panel a)) the data show the presence
of a broad mode centered near 4\,meV\@. At 300\,K (panel b)) this
mode softens considerably on cooling, extending all the way to the
elastic position, but it does so in such a way that it essentially
becomes a continuum of scattering over energy transfers from
0--5\,meV\@.

Second, another soft phonon is observed at the $R$-point
$\vec{Q}_R=(\frac{3}{2}, \frac{3}{2}, \frac{1}{2})$. At 300\,K there
is evidence of a column, but it is far weaker and much narrower in
momentum than that at $\vec{Q}_M=(\frac{3}{2}, \frac{3}{2}, 1)$.
This column appears to extend down to the elastic position clearly at 300 K. It
is also important to note the absence of any column at
$\vec{Q}_R=(\frac{3}{2}, \frac{3}{2}, \frac{3}{2})$.  We attribute
this to the phonon structure factor, and we use this observation
later when assigning the symmetries to the incipient soft modes at
$M$ and $R$\@.

A third feature, which is common to both the $M$- and $R$-points, is
that the strong softening of these zone-boundary modes has an
unusual dispersion that is localized in momentum but broad in
energy, which is what gives rise to the appearance of a column of
scattering in $\vec{Q}$-$E$ space. This feature is anomalous amongst
the ferroelectrics and perovskites, where soft modes are usually
described by a well-defined dispersion that is characterized by a
single, temperature-dependent value of energy for a given momentum
transfer, as seen near the $\Gamma$-point in several studies of PbTiO$_{3}$, PMN
and PMN-60PT\@.  Since the soft modes observed at both the $M$- and
$R$-points exhibit a continuum of energies at these specific
momentum transfers, they cannot be understood by such a well-defined
function.

\subsection{Triple-Axis Measurements at the $M$- and $R$-Points}

\begin{figure}[t]
\includegraphics[width=3.5in]{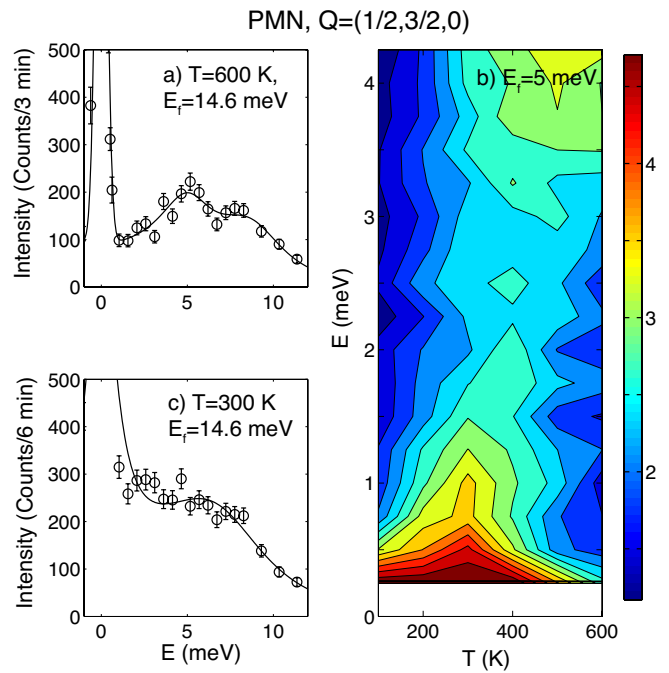}
\caption{Constant-$Q$ scans measured at the
$\vec{Q}_{M}=(\frac{1}{2}, \frac{3}{2}, 0)$ $M$-point using
triple-axis spectrometers.  Panels a) and c) show scans at 600\,K
and 300\,K, respectively, that demonstrate a dramatic softening on
cooling.  Panel b) is a color contour map, compiled from many
constant-$Q$ scans measured on SPINS, that illustrates the shift of
intensity between high and low energies as a function of temperature. The data were taken in the (HK0) scattering plane.}
\label{fig:m_constantQ}
\end{figure}

To determine the origin of the inelastic scattering, understand the
temperature dependence, and assign the soft mode to a particular
symmetry, we have conducted a series of triple-axis experiments in
the (HK0), (HHL), and (H3HL) scattering planes, each of which
provides access to different types of phonon modes.

We first examine the $M$-point phonons. Fig.~\ref{fig:m_constantQ}
displays two constant-$Q$ scans in which the scattered intensity was
measured at a fixed momentum transfer $\vec{Q}_{M}$=$(\frac{1}{2},
\frac{3}{2}, 0)$ while scanning the energy transfer ${\rm E} = \hbar
\omega = {\rm E}_i-\rm{E}_f$ at 600\,K (panel a)) and 300\,K (panel
c)). Whereas two well-defined modes are present at 600\,K, the 300\,K
data show a broad distribution of scattering resembling a continuum
that extends up to $\sim$7\,meV\@. The lineshape used to fit these
data is based on the sum of two damped, simple harmonic oscillators and has been
described elsewhere.~\cite{Stock05:74} Panel b) illustrates the
evolution of the phonon scattering with temperature at
$\vec{Q}_M =(\frac{1}{2}, \frac{3}{2}, 0)$ through a color contour
plot compiled from constant-$Q$ scans measured from 10 K to 600 K.  At high temperature little scattering is seen below
$\sim$4\,meV; at lower temperature (near 300\,K) a clear continuum
of scattering develops with intensity extending from 4\,meV to the
lowest energy transfer investigated.  The large intensity below
$\hbar\omega = 0.5$\,meV is attributed to pure elastic scattering
that is sampled due to the energy resolution of 0.25\,meV\@.
For temperatures below 300\,K, a decrease in the scattering is
observed in the energy range 0--4\,meV\@. This behavior is consistent
with that of the soft mode observed near the $\Gamma$-point and
illustrates the transfer of spectral weight from the inelastic
channel to the elastic channel, discussed in the previous section.

\begin{figure}[t]
\includegraphics[width=3.5in]{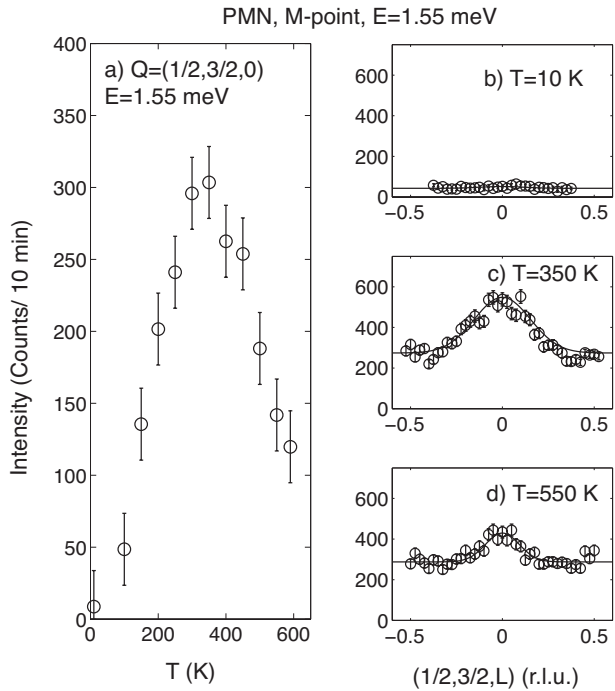}
\caption{A summary of constant-E scans measured at an energy $\hbar
\omega = 1.55$\,meV across the $\vec{Q}_{M}$=$(\frac{1}{2},
\frac{3}{2}, 0)$ $M$-point column.  Panel a) plots the intensity of
the peak observed in these scans as a function of temperature, and
representative scans are displayed at 10\,K, 350\,K, and 550\,K in
panels b), c), and d). The data were taken on the C5 (Chalk
River) thermal triple-axis spectrometer using E$_{f} = 14.6$\,meV\@ in the (H3HL) scattering plane.}
\label{fig:m_constantE}
\end{figure}

The temperature dependence of the $M$-point inelastic scattering was
characterized using constant-energy scans.  This method was chosen
over the conventional constant-$Q$ method because the column is a 
spectral feature that is well defined
in momentum, but broad in energy.  Therefore, a scan of momentum
transfer $\vec{Q}$, at a fixed energy $E$, will result in a well-defined
peak centered at the zone boundary.

Fig.~\ref{fig:m_constantE} illustrates constant energy scans
measured at $\hbar\omega = 1.55$\,meV over 10 K to 600\,K\@.  The data
were taken on the C5 triple-axis spectrometer using a fixed final
neutron energy $\rm E_f = 14.6$\,meV\@. Panel a) summarizes the peak
intensity, taken as the amplitude of a Gaussian fit to the
correlated scattering in momentum across the $M$-point, as a
function of temperature. On cooling, the intensity at
1.55\,meV increases, peaks at $T \sim 400$\,K, and rapidly declines,
becoming nearly unobservable by 10\,K\@. Examples of the fits at
several temperatures are illustrated in panels b)-d).

\begin{figure}[t]
\includegraphics[width=3.5in]{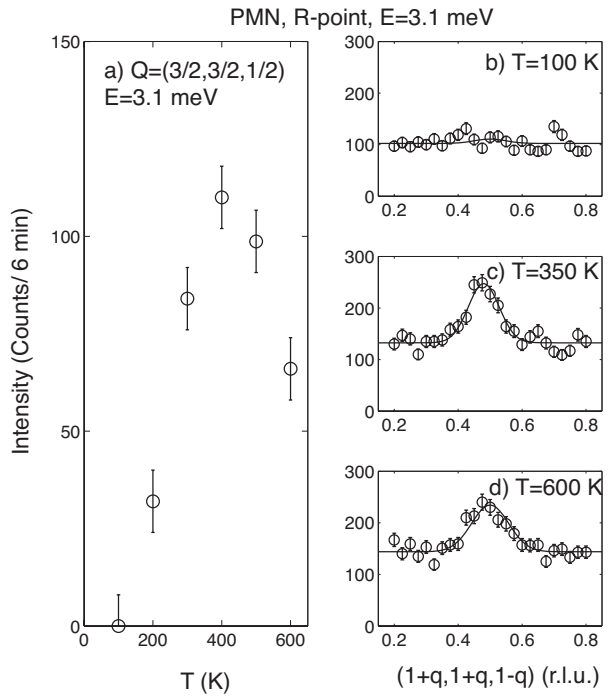}
\caption{A summary of constant-E scans conducted at $\hbar\omega =
3.1$\,meV through the $R$-point column at
$\vec{Q}_{R}$=$(\frac{3}{2}, \frac{3}{2}, \frac{1}{2})$.  Panel a)
plots the peak intensity observed in scans through the inelastic
column, and representative scans are displayed at 10\,K, 350\,K, and
600\,K in panels b), c), and d).  The data were taken on the N5
(Chalk River) thermal triple-axis spectrometer with fixed $E_f =
14.8$\,meV\@ in the (HHL) scattering plane.} \label{fig:r_constantE}
\end{figure}

Whereas our preceding discussion has focused around the $M$-point, the
$\vec{Q}$-$E$ data taken on DCS reveal a second column at the
$R$-point.  Fig.~\ref{fig:r_constantE} shows constant-energy scans
at $\hbar \omega = 3.1$\,meV\@. Panel a) shows the temperature
dependence of the column.  Panels b)--d) display representative
constant-energy scans fit to Gaussian peaks centered at $R$\@. The
scattering at $R$ (Fig.~\ref{fig:r_constantE})
shows a temperature dependence on
cooling nearly identical to that observed at $M$ (Fig.~\ref{fig:m_constantE}),
reaching a maximum intensity at 400\,K and declining at lower
temperatures.  The temperature dependence of the scattering observed
at both $R$ and $M$ cannot be accounted for by the Bose factor alone; 
if this were the case, then the intensity at a fixed energy
transfer would continue to grow with increasing temperature.  
Therefore, our results show that there are incipient soft
modes at both $M$ and $R$ that reach a minimum frequency (hence
giving maximum intensity along the energy column) at $\sim$400\,K\@.

\begin{figure}[t]
\includegraphics[width=2.75in]{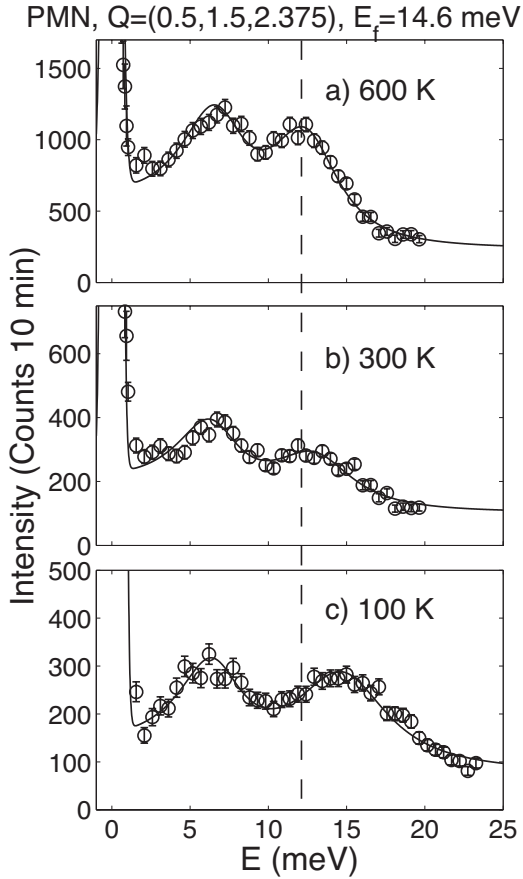}
\caption{Constant-$Q$ scans conducted at a wave vector displaced from
the $R$-point at 600\,K, 300\,K, and 100\,K\@.  The data illustrate a
softening of the optic mode that recovers to an energy of
$\sim$15\,meV at 100\,K\@. The data were taken at the C5 (Chalk River)
thermal triple-axis spectrometer with fixed E$_f = 14.6$\,meV\@.}
\label{fig:r_constantQ2}
\end{figure}

\begin{figure}[t]
\includegraphics[width=3.0in]{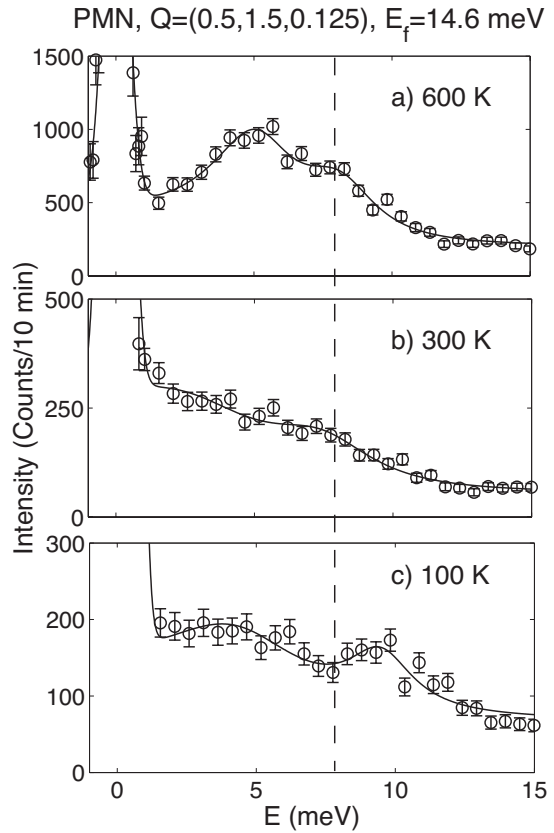}
\caption{Constant-$Q$ scans conducted at a wave vector displaced from
the $M$-point at 600\,K, 300\,K, and 100\,K\@.  The data illustrate a
softening of the optic mode, which recovers to an energy of 
$\sim$10\,meV at 100\,K\@.  The data were taken at the C5 (Chalk River)
thermal triple-axis spectrometer with fixed E$_f = 14.6$\,meV\@.}
\label{fig:m_constantQ2}
\end{figure}

It is important to work out whether the soft modes at $M$ and $R$
are acoustic (tilts of the octahedra) or optic modes.  It is difficult to answer this
question using scans exactly at these high-symmetry points, where
the soft modes form a broad continuum in energy.  We therefore found
it useful to conduct constant-$Q$ scans at wave vectors slightly
displaced from $M$ and $R$, where the acoustic and optic modes could
be resolved at all temperatures.  

Constant-$Q$ scans near the $R$-point are presented in
Fig.~\ref{fig:r_constantQ2}, where the solid line is the result of a
fit to two damped harmonic oscillators.  Two distinct modes are
observed, with the lower-energy mode showing little temperature
dependence while the higher-energy mode hardens upon cooling to 100 K.  As
this upper mode is temperature dependent, we associate the
columns of scattering at the $R$-point with this optic mode, and not the tilts. 
This point is confirmed directly using structure factors and group theory analysis in the final section of this paper. We have not pursued the temperature dependence of the frequencies at high temperatures (to observe the softening) due to the possibility of sample breakdown.  This softening would require measurements over a very broad temperature range as the rate of softening with temperature is predicted, and has been observed in conventional ferroelectrics, to be much less than the rate of recovery at low temperatures.  

Similar scans were also conducted near $M$ and are presented
in Fig.~\ref{fig:m_constantQ2}. The solid line in the figures is a fit to
the sum of two damped harmonic oscillators. For scans at $\vec{Q} =
(\frac{1}{2}, \frac{3}{2}, 0.125)$, which lies close to $M$, two
low-energy phonon modes could be clearly observed at 600\,K (panel
a) of Fig.~\ref{fig:m_constantQ2}).  At 300\,K (panel b)) only a
continuum of scattering is observed, but a recovery of a higher energy branch is apparent at
100\,K (panel c)), where the higher energy mode recovers to an
energy transfer of $\hbar \omega = 10$\,meV\@.  Whereas the interpretation of the low-energy scattering in these scans is ambiguous, because the $M$-column is intense and distributed in momentum (see Fig.\ref{fig:DCS}) and there are several modes that may be close in energy (see Ref.~\onlinecite{Boyer81:24}), these scans do show that there exists a higher-energy mode that is temperature dependent.  This implies that an optic mode contributes strongly to the column of scattering at this wave vector.  We rule out the contribution from zone-boundary acoustic, or tilt, modes through a discussion of structure factors and group theory presented later in the paper.

\subsection{Comparison with a ``Conventional'' Ferroelectric (PMN-60PT)}

We now make a brief comparison with PMN-60PT, a conventional
ferroelectric that undergoes a first-order, cubic--tetragonal
transition at 540\,K\@.  The large concentration of PT places PMN-60PT
beyond the MPB, which divides relaxor ferroelectrics, defined by
broad temperature- and frequency-dependent peaks in the dielectric
susceptibility, from conventional ferroelectrics, defined by sharp
peaks in the dielectric susceptibility and well-defined structural
transitions. The transition in PMN-60PT is driven by a soft TO~mode and has been
studied as a function of temperature.  The soft TO~mode is the same
mode that softens in relaxors, such as PMN and PZN, and that possibly
gives rise to the static diffuse scattering.

\begin{figure}[t]
\includegraphics[width=3.5in]{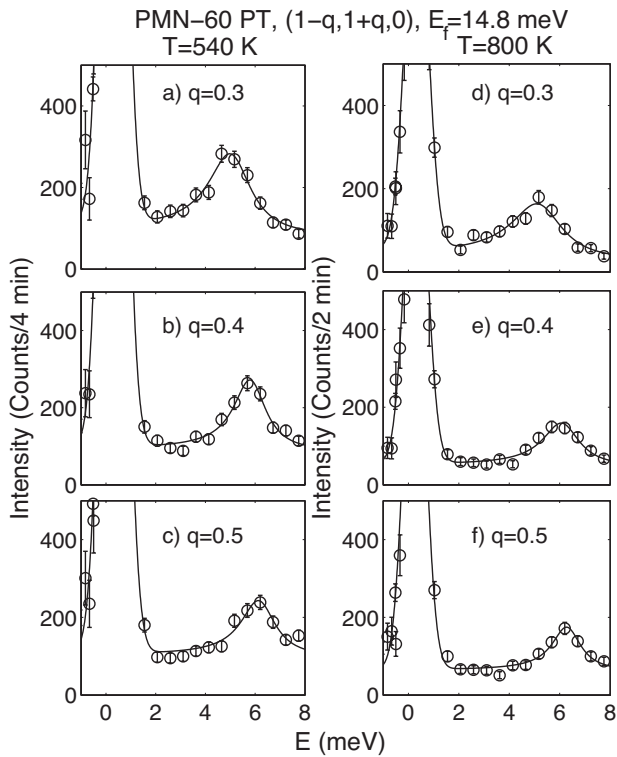}
\caption{Constant-$Q$ scans measured near the $M$-point in PMN-60PT
at 800\,K and at 540\,K, in the vicinity of the cubic--tetragonal
structural transition.  The data were taken on the N5 (Chalk River)
spectrometer with fixed E$_f = 14.8$\,meV\@.} \label{fig:PMN_60PT}
\end{figure}

A summary of the phonon dispersion, measured in the (HK0) scattering plane,
is illustrated in Fig.~\ref{fig:PMN_60PT} for $T=540$\,K (panels a)-c))
and $T=800$\,K (panels d)-f)) near $M$\@.  At the zone boundary
(panels c) and f)), a well-defined phonon mode is observed
near 6\,meV at both 800\,K and 540\,K\@. The phonon linewidth is smaller than the energy of the
phonon and therefore the peaks represent
\textit{underdamped} propagating phonon modes.  The phonon energy is
unchanged within error on cooling from 800\,K to 540\,K, a temperature very near 
to the structural transition temperature.  The phonon
peaks are well defined and no column or continuum of scattering is
observed near $M$ ($q=0.5$).  Constant-$Q$ scans at $q=0.3$ (panels
a) and d)) and at $q=0.4$ (panels b) and e)) show that, in contrast
to PMN, well-defined phonon modes are observed. No
continuum scattering has been reported in pure PbTiO$_{3}$ (PT) and
we therefore assume it to be absent as it is in
PMN-60PT\@.~\cite{Tomeno06:73}

It seems likely that the soft columns at the high-symmetry points of
the Brillouin zone edge are only present in the relaxor phase and
possibly only for PT concentrations below the MPB\@.  This
will need to be verified through a study of several other lead-based
relaxor ferroelectrics, particularly near the MPB concentration,
which for PMN-$x$PT is $x =32\%$.

\section{Group Analysis}

Having identified that the columns at $R$ and $M$ are the result of
soft optic (not acoustic or tilt) modes, we now investigate which atoms
contribute to these modes through the use of group theory and
structure factors. The history of the lattice
dynamical studies of perovskites is so long that various systems have been
used to label the eigenmodes of the phonons, including a system used
by Cowley,~\cite{Cowley64:134} Miller and Love,~\cite{Miller-Love}
and Bouckaert, Smoluchowski, and Wigner (BSW)\@.~\cite{BSW}  The
translation between the labeling system of Cowley and BSW is given
by Boyer\@.~\cite{Boyer81:24} The established setting of the crystal
in most lattice dynamical studies of perovskites is one where the
octahedron sits in the center of the unit cell, centered on the
Wyckoff $b$-sites with O on the $c$-sites, and Pb on the corner
$a$-sites (Fig.~\ref{fig:struct-BZ} a). This contrasts with most
discussions of the structure and group theory underlying the tilt
transitions in perovskites, in which, the octahedron conventionally
sits on the corner of the unit cell.~\cite{Howard-Stokes98,Howard-Stokes02}  This difference in the cell
setting causes the mode labels, even within the same labeling
system, to change. In order to make a comparison to the majority of
the lattice dynamics studies we use the former setting, typically
used for dynamical studies. The decompositions were performed with
SMODES,\cite{Isotropy} and the compatibility relations were derived
with the Isotropy~\cite{Isotropy} code.  We calculated the structure
factors based on the mode eigenvectors derived from
Isotropy~\cite{Isotropy} and visualized them with
Isodisplace.~\cite{Isodisplace}  We use the labels of Miller and
Love.~\cite{Miller-Love}

\begin{figure}[t]
\includegraphics[width=3.0in]{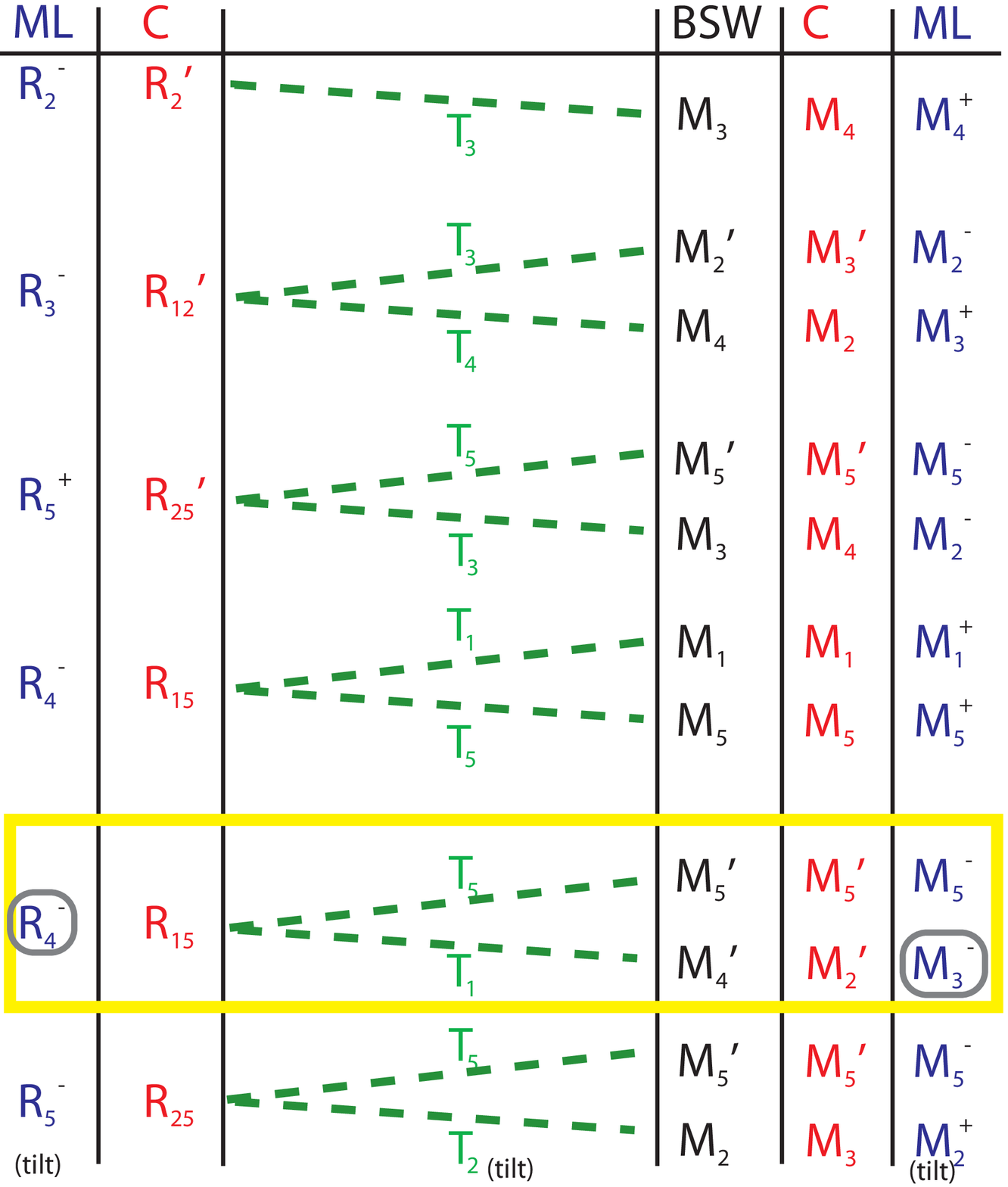}
\caption{ The compatibility of the irreducible representations of
phonon modes propagating along $M$-$T$-$R$ is illustrated.  The two
lowest modes ($R_4^-$ and $R_5^-$) are the lowest energy modes in
SrTiO$_3$,~\cite{Cowley64:134} RbCaF$_3$,~\cite{Boyer81:24}
CaSiO$_3$,~\cite{Stixrude96:81} KNbO$_3$,~\cite{Yu95:74}
PbTiO$_3$,~\cite{Garcia96:54,Tomeno06:73}
BaTiO$_3$,\cite{Shirane70:2} KTaO$_3$,~\cite{Perry89:39}, and
PbZrO$_3$\@.~\cite{Ghosez99} ML stands for the labeling system of
Miller and Love~\cite{Miller-Love}, C for that of
Cowley~\cite{Cowley64:134} and BSW that of Bouckaert, Smoluchowski,
and Wigner\@.~\cite{BSW}  The symmetries of the zone boundary columns are highlighted.} \label{fig:group_theory}
\end{figure}

\begin{figure}[t]
\includegraphics[width=3.5in]{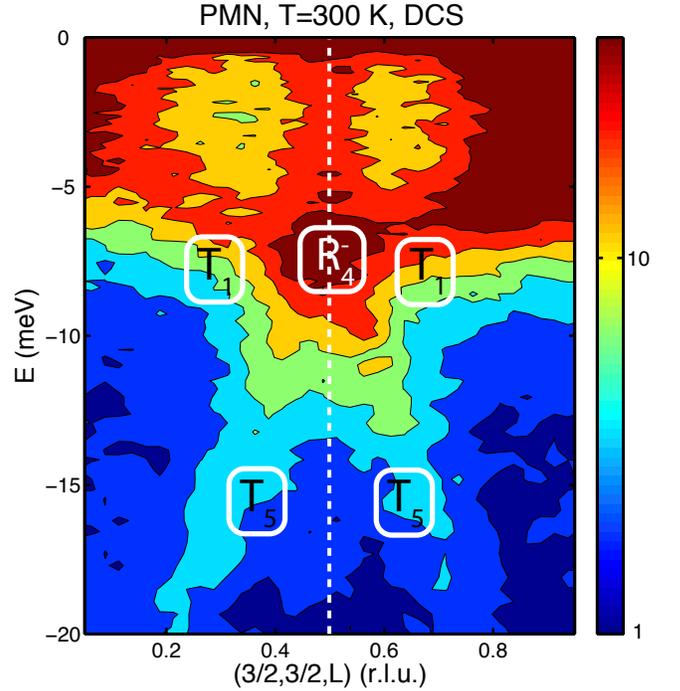}
\caption{Intensity contours of the phonon scattering near the
$R$-point measured at $T=300$\,K, plotted on a logarithmic scale. A higher-energy 
mode originating from the $R$-point (labelled T$_{5}$) is clearly seen extending
up to at least $\sim$20\,meV energy transfer.  The data were taken
on DCS with E$_i$=10.4\,meV\@.} \label{fig:R_disperson}
\end{figure}

\subsection{Mode Decomposition and Structure Factor Analysis}

The Brillouin zone is shown in Fig.~\ref{fig:struct-BZ}b, along with
labels of high-symmetry points and lines, using the system of Miller
and Love\@.~\cite{Miller-Love} The total decomposition of the atomic
motions into irreducible representations was calculated at $\Gamma$,
$M$, and $R$ and compared to that derived by
Cowley\@.~\cite{Cowley64:134} A compatibility diagram is shown in
Fig.~\ref{fig:group_theory}, which also relates our notation to
those listed previously. To remove ambiguity as to which modes at
$M$ are linked to those at $R$, published, measured, and calculated
dispersion curves of SrTiO$_3$,~\cite{Cowley64:134}
RbCaF$_3$,~\cite{Boyer81:24} CaSiO$_3$,~\cite{Stixrude96:81}
KNbO$_3$,~\cite{Yu95:74} PbTiO$_3$,~\cite{Garcia96:54,Tomeno06:73}
BaTiO$_3$,~\cite{Shirane70:2} KTaO$_3$,~\cite{Perry89:39} KMnF$_{3}$~\cite{Gesi72:5} and
PbZrO$_3$~\cite{Ghosez99} were consulted.  There are relatively few
$R$-modes. A common feature of the low-frequency lattice dynamics of
all the perovskites listed above is that the two lowest frequency
$R$-modes are $R_{5}^{-}$ and $R_{4}^{-}$, whereas the highest
frequency mode is $R_2^{-}$. Owing to the number of modes at $M$, there
are fewer general observations that can be made about the spectra at this wave vector.

The $R$-point is the corner of the Brillouin zone,
which, being a cube, lies at the convergence of three $T$~zone edges
(Fig.~\ref{fig:struct-BZ}b). Therefore, in general, the $R$-point
irreducible representations are of higher dimensionality than those
at $M$; on moving from $R$ to $T$ the loss of degeneracy means that,
with the sole exception of the highest-frequency $R_2^{-}$ mode,
each $R$-mode splits into two $T$-branches terminating in a single
$M$-mode\@. Fig.~\ref{fig:R_disperson} plots contours of phonon
intensity of the column near the $R$-point at 300\,K, based on data
taken on the DCS instrument.  The $x$-axis is the $\vec{Q}
=(\frac{3}{2},\frac{3}{2},{\rm L})$ direction, i.e., the $T$-line
denoted in Fig.~\ref{fig:struct-BZ}b.  The plot shows the soft mode
at $\vec{Q}_R =(\frac{3}{2},\frac{3}{2},\frac{1}{2})$ splitting into
the two $T$-branches: one low-energy mode that resides between
5--8\,meV, and another mode that rises rapidly in energy to over
20\,meV\@.

One famous feature of the low-frequency dynamics of perovskites is
the tilt modes.~\cite{Swainson05:61} All of the
commensurate tilt structures of perovskites can be related to
combinations of various dimensions of two zone-boundary acoustic
modes at $M$ and $R$\@.~\cite{Howard-Stokes98, Howard-Stokes02} The
$T$-line that joins these points has also been shown to have tilt
modes,~\cite{GiddyA49:93} and the whole $M$-$T$-$R$ line has been
observed to soften in some cases such as
SrTiO$_3$\@.~\cite{Stirling72:5} The tilt mode at $R$ is $R_5^-$,
which splits into one low-frequency $T_2$~tilt mode, terminating in
the $M_2^+$~tilt mode, and one non-tilt $T_5$~mode, which rises
rapidly in frequency, terminating in a mode of symmetry $M_5^-$
(Fig.~\ref{fig:group_theory}). Similarly, the low-frequency
$R_4^-$~optic mode splits into the $T_5$- and $T_1$-branches, which
terminate in $M_5^-$ and $M_3^-$, respectively: 
there is, therefore, the potential
for an anti-crossing (Fig.~\ref{fig:group_theory}) of the
$T_5$-branches in the low-frequency dynamics, as is observed near $R$
in RbCaF$_3$~\cite{Boyer81:24} and CaSiO$_3$\@.~\cite{Stixrude96:81}
These represent the likeliest candidates to be observed in the
lowest frequency part of the spectra. To avoid relying solely on these 
general observations, we have examined the structure factors of all the $M$ and $R$ modes so as
not to exclude other possibilities.

\begin{figure}[t]
\includegraphics[width=2.8in]{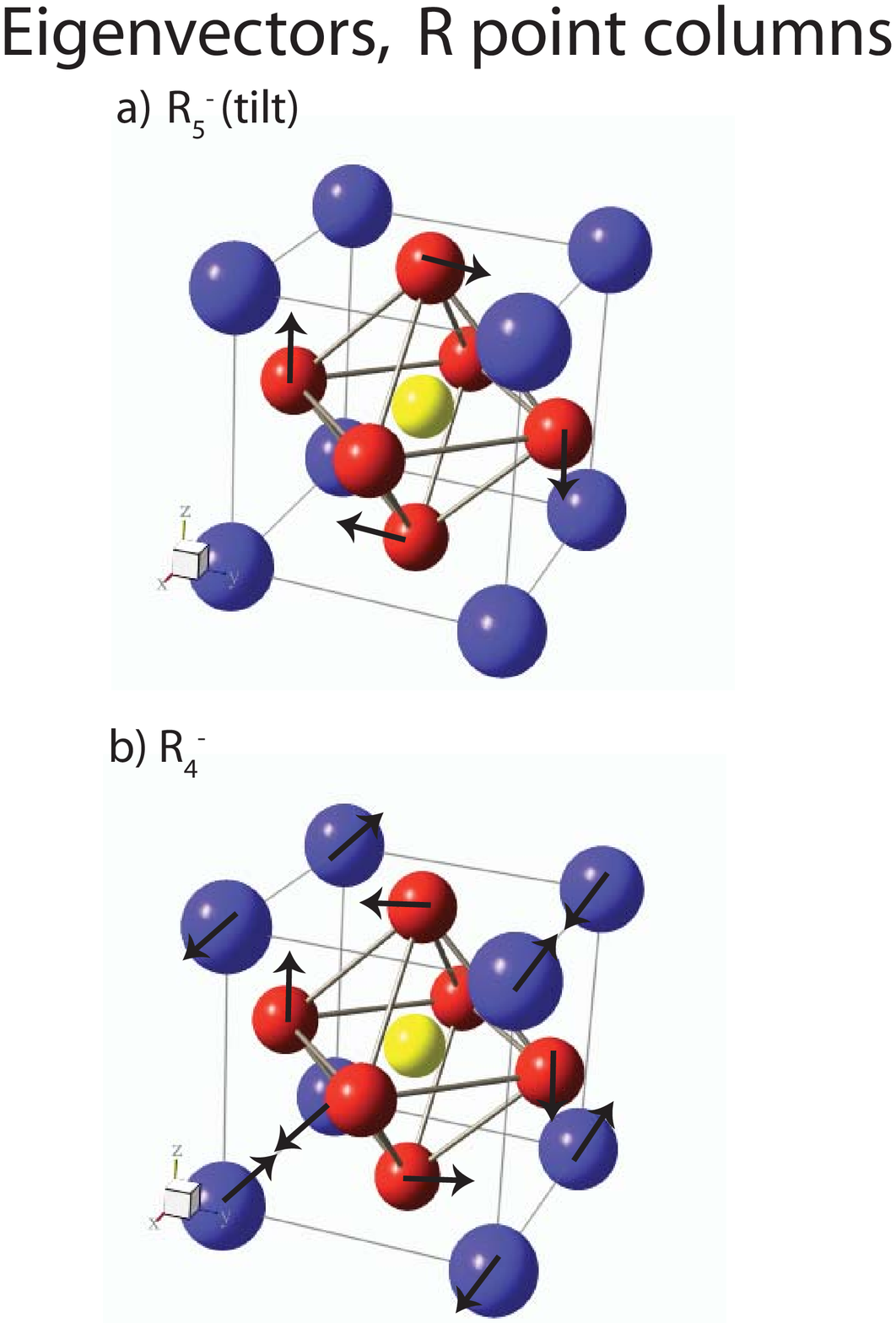}
\caption{The possible eigenmodes resulting in the $R$-point zone
boundary column. The blue, red, and yellow atoms correspond to
Pb$^{2+}$, O$^{2-}$, and Mg$^{2+}$/Nb$^{5+}$ ions, respectively.
Panel a) illustrates the tilt displacements about the $x$-axis
corresponding to a single component of $R_5^-$; all neighboring
octahedra above and below would be tilted out-of-phase.  Panel b)
illustrates the displacements of a single-$k$ distortion of the
$R_4^-$ optic mode. Based on a comparison with SrTiO$_3$ and a
structure factor analysis, we suggest that the column is associated
with displacements of the form shown in b), i.e., $R_4^-$\@.  }
\label{fig:eigenmodes}
\end{figure}

The inelastic structure factor is given by the following formula for
the \textit{m}-th mode,

\begin{eqnarray}
{F_m(\vec Q)}={\sum\limits_{n} \left({\vec{Q}} \cdot {{\vec{\xi}}_{m,n}\over m_n}\right) b_n e^{i{\vec{Q}}\cdot{\vec{r_n}}}},
\end{eqnarray}

\noindent where $\vec{Q}$ is the total momentum transfer, 
$\xi_{m,n}$ is the eigenvector
for the $m$-th eigenmode,  and
$r_n$ is the position, $b_n$ the
scattering length, and $m_n$ the mass
of the $n$-th atom.  In this equation $n$ sums over all atoms in the
unit cell.\cite{Harada70:26}  Since we are working at a fixed
temperature and for comparably small momentum transfers, we have
ignored the Debye-Waller factor.  Using this formula and the
eigenvectors obtained from Isotropy,\cite{Isotropy} we find that
tilt distortions at $\vec{Q}_R
=(\frac{3}{2},\frac{3}{2},\frac{1}{2})$, where we observe the strong
column of scattering, have identically zero structure factor in the
zones in which the DCS spectra were measured.  Therefore, the
modes contributing to the columns must
instead be zone-boundary optics, i.e., dynamic antiferroelectric
distortions.

By comparison with many other cubic perovskites, the likeliest next-lowest
frequency optic mode at $R$ is $R_4^-$ (Fig.~\ref{fig:group_theory}).
This is particularly clear in the calculations of Garcia and
Vanderbilt~\cite{Garcia96:54} for PbTiO$_3$ that have been confirmed
experimentally by Tomeno~\textit{et al.}~\cite{Tomeno06:73} In
addition to the general observation about the order in frequency of
modes at $R$, calculations of the structure factors of $R$-point
phonons show that a single-$k$ structure of $R_4^-$, in which Pb and
O are displaced, can lead to near-zero neutron structure factors at those
$R$-points for which all indices are equal, as observed in our scattering data at $(\frac{1}{2},\frac{1}{2},\frac{1}{2})$,
$(\frac{3}{2},\frac{3}{2},\frac{3}{2})$, etc.

Having assigned the $R$-column based on a combination of group
theory and structure factor analysis, and having
eliminated the tilt modes from consideration, we now assign the
displacements associated with the $M$-column\@.  As illustrated in
Fig.~\ref{fig:m_constantQ2}, a single $T$-branch appears
to be interrupted by columns at the high-symmetry $M$- and
$R$-points\@. As the $T_5$-branch, due to anti-crossing, typically
rises rapidly in energy in other perovskites, the symmetry of the $M$-column is
probably compatible with the $T_1$-branch. Assuming this, we assign the symmetry of the
$M$-column phonon to $M_3^-$\@. This conclusion is corroborated by
structure factor calculations. We observe a soft
phonon column at $M$-points of all forms, including
$(\frac{3}{2},\frac{3}{2},0)$ and $(\frac{1}{2},\frac{3}{2},0)$. The
inelastic structure factor of the $M_3^-$ phonon is finite at all
$M$~positions and grows with increasing $|\vec{Q}|$\@.   The
eigenvectors of $M_3^-$ consist only of displacements of Pb~ions
along the $\langle 100\rangle$~directions.  We note that this conclusion is in agreement with calculations presented in Ref.~\onlinecite{Prosandeev04:70}, which predict a low energy phonon mode consisting of antiphase  displacements of Pb\@. 

\section{Discussion}

It is interesting to compare the lattice dynamics of relaxors to
those of conventional ferroelectrics.  Two notable features are
present in PMN\@.  The first is the ``waterfall'' effect, which is
characterized by a broadening in energy of the low-energy TO~mode
for wave vectors near the zone center.  This phonon anomaly is also
present in PMN-60PT, which exhibits a sharp structural phase
transition, and in relaxor ferroelectrics having PT concentrations
below the critical concentration of 32$\%$, which defines the MPB\@.
As the exceptional dielectric properties of relaxor ferroelectrics
are absent in PMN-60PT, the ``waterfall'' effect cannot be linked to
relaxor behavior.  The second notable feature is the presence of columns of
inelastic scattering observed at the $M$- and $R$-point zone
boundaries in PMN\@. These columns exhibit a similar softening near
400\,K, where strong diffuse scattering is onset (as shown in other
experiments~\cite{Hiraka04:70,Gehring_TB}), but they are not present
in PMN-60PT\@.  Therefore, whereas the zone-boundary soft phonons may
not be the direct cause of the unique dielectric properties in PMN,
they may be indirectly coupled and indicative of the underlying
relaxor ferroelectricity.

There is no obvious physical reason for these columns of scattering.
Similar columns have been observed in several other systems;
examples include TTF-TCNQ (Ref.~\onlinecite{Shirane76:14}),
K$_2$Pt(CN)$_4$Br$_{0.3}$ $\cdot$ 2.3 D$_2$O (Ref.~\onlinecite{Shapiro76:14,Carneiro76:13}), 
K$_{0.3}$MoO$_3$ (Ref.~\onlinecite{Pouget91:43}), 
and body-centered cubic zirconium (Ref.~\onlinecite{Heiming91:43})\@.  
In the above cases, the near vertical rod of
phonon scattering has been attributed to a Kohn anomaly or a
coupling between electronic and lattice degrees of
freedom.~\cite{Cowley06:75}  Columns of scattering in insulators,
where no such coupling can exist, are much rarer but have been
observed in CaCO$_3$ (Ref.~\onlinecite{Harris98:10}) and KCN (Ref.~\onlinecite{Rowe78:40}). 
Both of these systems display molecular
rotations and the transitions are order-disorder in nature; here the
columns were interpreted in terms of a coupling between molecular
relaxational modes and harmonic lattice vibrations. Theories
describing such coupling have been proposed by Michel and Naudts
(Ref.~\onlinecite{Michel78:68}) and similar models have been used by
Shapiro \textit{et al.} (Ref.~\onlinecite{Shapiro72:6}) to describe
the central peak in SrTiO$_3$\@.  Therefore,
columns of scattering have been associated with a coupling of the
lattice (harmonic degrees of freedom) with another degree of freedom
(either electronic or molecular for the examples cited).

The soft phonon columns in PMN are therefore very unusual.  The
lineshape, which is localized in momentum but broad in energy,
cannot be associated with an electronic process (such as a Kohn
anomaly) as PMN is an insulator and therefore there is no Fermi
surface.  Nor can they be directly associated with a coupling to
molecular rotations, as no such local molecular ions exist in the PMN structure. One possibility is of a local rattling of atoms, 
which is directly related to the disorder in PMN\@. The
local disorder and random fields caused by the charge difference
between Nb$^{5+}$ and Mg$^{2+}$ may introduce a number of local
states that can be occupied (in analogy to molecular solids), to
which the lone pair displacement of Pb$^{2+}$ ions can couple.  Such
structures have been proposed to understand powder diffraction
measurements in PZN (Ref.~\onlinecite{Iwase99:60}) and PMN 
(Ref.~\onlinecite{Bonneau89:24}) as well as NMR measurements (Ref.~\onlinecite{Blinc03:91})\@. 
The role of Pb$^{2+}$ ion displacements in the
unusual piezoelectric properties was also proposed based on PDF
analysis.~\cite{Egami98:32} Therefore, as the same atoms can be on
several sites in different unit cells, this may act as a means for a
relaxational mode that can then couple to the harmonic modes at high
energies.  Columns are not observed in
PMN-60PT, which lacks any temperature dependent zone-center or 
zone-boundary elastic diffuse scattering, but are present in PMN, where
strong zone-boundary elastic diffuse scattering is present, inferring
that static, short-range correlations are required for columns to
form.  The columns of scattering seem to represent a series of
nearly degenerate states, which must be related to the formation of
a non-zero polarization, as evidenced by the soft mode at the
$\Gamma$-point. The presence of several nearly degenerate
states is also evident from the complex structural phase diagrams
and properties observed as a function of PT-doping\@.~\cite{Cao06:73}  We note
that this idea is completely heuristic, and thus it will be
interesting if future calculations and theories can reproduce these
columns.

The temperature dependence of the soft phonon columns and the fact that the softening occurs at nearly the same temperature as the minimum in the frequency of the zone-center TO~mode and the onset of the diffuse scattering is strongly suggestive that the zone-boundary columns maybe a signature of the relaxor state.  This is confirmed by the absence of of these columns in PMN-60PT, a well-defined ferroelectric.  The only conclusive means of proving the direct connection with the relaxor state is a detailed study as a function of PT-concentration.

Our results point to an underlying antiferroelectric distortion in the relaxor PMN that competes with the ferroelectric distortion, characterized by strong, zone-center diffuse scattering and a soft optic mode.  The presence of strong random fields and competing interactions provides an underlying physical reason for the glassy nature observed using dielectric measurements.  How these competing interactions change with doping and, in particular, near the MPB will be a topic of interesting future study. 

\section{Conclusion}

We have discovered the existence of very unusual soft modes in PMN
at the $\vec{Q}_R=\left\{\frac{1}{2},\frac{1}{2},\frac{1}{2}\right\}$ and
$\vec{Q}_M=\left\{\frac{1}{2},\frac{1}{2},0\right\}$ zone boundaries.  The phonon modes have an anomalous dispersion, are localized in momentum,  but broad
in energy resulting in a column of scattering.  The columns of
scatter is present in PMN (a relaxor ferroelectric) are absent
in PMN-60PT (a well-defined ferroelectric).  The physical origin of
the unusual dispersion is not understood and cannot be accounted for
by a coupling between electronic and lattice degrees of freedom, nor
the presence of a molecular relaxational mode.  The
prevalence of many inequivalent local atomic sites may result in a
relaxational mode, in analogy to order-disorder systems.  It is possible that 
a strong coupling between zone-center and
zone-boundary distortions 
may be a general component of relaxor ferroelectrics, which requires verification for further compositions on either side of the MPB\@.

\begin{acknowledgments}

We thank R. Cowley, M. Gutmann, and J. Lynn for helpful discussions and B. Clow, T. Dodd, L. McEwan, R. Sammon, M. Watson and T. Whan for technical support during experiments.  We acknowledge financial support from the Natural Science and Engineering Research Council of Canada (NSERC), DMR-9986442, and from the U.S. DOE under contract No. DE-AC02-98CH10886, and the Office of Naval Research under Grant No. N00014-99-1-0738.  This work utilized facilities supported in part by the National Science Foundation under Agreement No. DMR-045467. Part of this work was also funded under the Graduate Supplement Scholarship program from the National Research Council (NRC) of Canada.

\end{acknowledgments}


\end{document}